\documentclass[%
 reprint,
 amsmath,amssymb,
 aps,
pra,
]{revtex4-1}
\usepackage{silence}
\usepackage[utf8]{inputenc}
\usepackage{graphicx}
\usepackage{dcolumn}
\usepackage{bm}
\usepackage{ulem}
\usepackage{xcolor}
\usepackage{lipsum}
\usepackage{hyperref}
\hypersetup{
    colorlinks=true,
    linkcolor=blue,
    filecolor=magenta,      
    urlcolor=black,
    citecolor=blue,
    pdftitle={Overleaf Example},
    pdfpagemode=FullScreen,
    }

\begin{document}

\preprint{APS/123-QED}


\title{Quantum-enhanced phase sensitivity in an all-fiber Mach–Zehnder interferometer}

\author{Romain Dalidet$^1$, Sébastien Tanzilli$^1$, Gregory Sauder$^1$, Nicolas Fabre$^2$, Laurent Labonté$^1$}
\email{laurent.labonte@univ-cotedazur.fr}
\author{Anthony Martin$^1$}

\affiliation{$^1$Université Côte d’Azur, CNRS, Institut de physique de Nice, France \\ 
$^2$Télécom Paris-LTCI, Institut Polytechnique de Paris, 19 Place Marguerite Perey, 91120 Palaiseau, France}

\begin{abstract}
Recent advances in quantum photonics have enabled increasingly robust protocols in optical phase estimation, achieving precisions beyond the standard quantum limit and approaching the Heisenberg limit. While intrinsic losses hinder the realization of unconditional super-sensitivity, reaching quantum advantage, defined as sensitivity surpassing that of any classical counterpart with identical resources, remains achievable. Here we experimentally demonstrate such an advantage using a fully fibered Mach–Zehnder–type interferometer operating at telecom wavelengths, free of post-selection. The scheme relies on the conversion of polarization-entangled photon pairs, a degree of freedom commonly favored for experimental convenience, into energy–time entanglement, which is particularly well suited for scalable fiber-based sensors. All system imperfections, including asymmetric losses and detector inefficiencies, are accounted for in the Fisher information analysis, yielding a measured quantum advantage of 10\%. This result highlights the practicality of compact, alignment-free quantum interferometers for real-world sensing applications. 
 \end{abstract}

\maketitle

\section{Introduction}

Achieving ultra-precise measurements is a central goal across science and engineering, ranging from probing fundamental physical constants to enhancing sensing technologies \cite{Giovannetti2011}. In optical interferometry, the ultimate sensitivity to phase shifts is constrained by the quantum fluctuations of light \cite{Helstrom1969, Braunstein1994, Giovannetti2004}. By exploiting non-classical states, such as squeezed or entangled states, quantum metrology can overcome the standard quantum limit (SQL) and, in principle, approach the Heisenberg limit, where sensitivity scales inversely with the total photon number \cite{Giovannetti2006, Polino2020}.
In practice, however, intrinsic losses, detector inefficiencies, and other experimental imperfections severely limit the Fisher information (FI) that can be extracted per photon, making unconditional super-sensitivity extremely challenging ~\cite{DEMKOWICZDOBRZANSKI2015345}. While surpassing the SQL under realistic conditions is notoriously difficult~\cite{PhysRevLett.123.231107}, quantum advantage offers a relaxed, yet genuine quantum benchmark, capturing any performance gain over the optimal classical strategy with identical energy resources under realistic imperfections.

While polarization-entangled photon pairs have enabled landmark demonstrations of quantum-enhanced interferometry \cite{Slussarenko2017, PhysRevLett.130.070801}, they are considered to be less suited to realistic sensing scenarios in optical fibers. More specifically, polarization is susceptible to uncontrolled rotations and drifts in deployed networks, and most practical measurands such as temperature, strain, or rotation—affect the phase accumulated along the fiber rather than its birefringence. Many sub–shot-noise measurements that do not use polarization have also relied on bulk optical setups, limiting scalability and compatibility with fiber-based, phase-sensitive sensing \cite{Slussarenko2017, PhysRevLett.130.070801, PhysRevLett.119.223604,PhysRevLett.128.033601}. In contrast, energy–time (ET) entanglement is inherently compatible with fiber transmission and naturally probes phase shifts arising in the optical path, making it an ideal resource for flexible, alignment-free quantum sensors operating outside the laboratory.

To date, most experimental demonstrations with N00N states neglect realistic imperfections in order to assess high measurement sensitivity \cite{10.1063/1.4724105, Fink2019}. Here, we report the first demonstration of genuine quantum advantage in a fully fibered Mach–Zehnder–type interferometer at telecom wavelengths. Our approach transcribes polarization entangled photon pairs into ET entanglement within a folded Franson interferometer, enabling deterministic photon separation and full FI reconstruction. All relevant system imperfections, including asymmetric losses and detector efficiencies, are included in the analysis, yielding a measured quantum advantage of 10\%. This result establishes energy–time entanglement as a viable resource for real-world quantum-enhanced sensing.

\section{Experimental strategy}

\begin{figure*}[!t]
    \centering
    \includegraphics[width=1\linewidth]{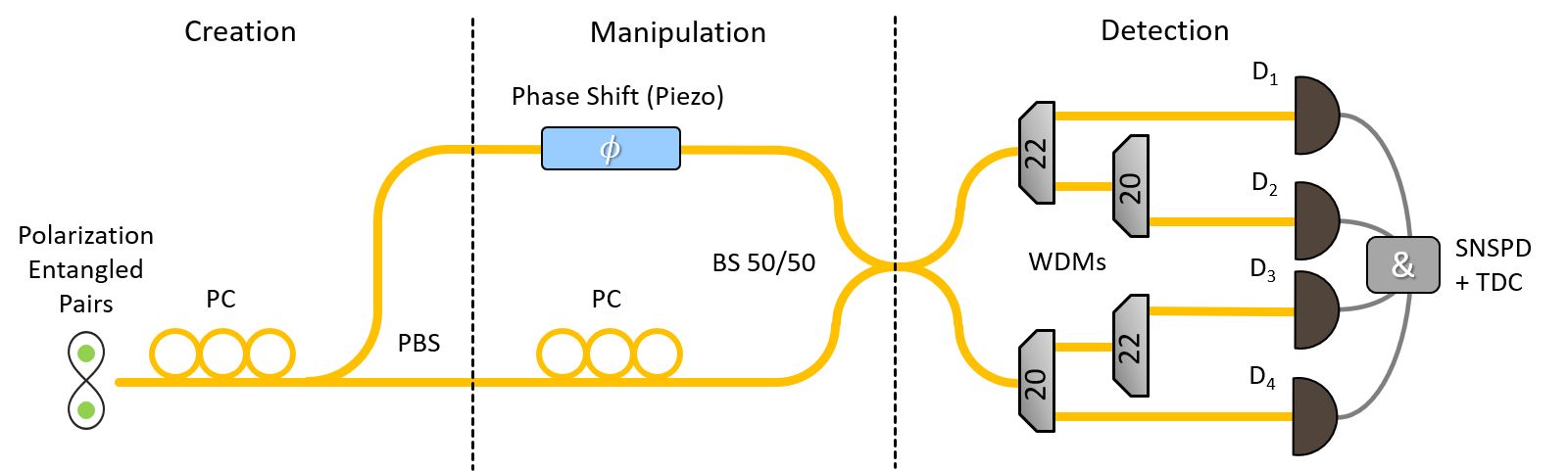}
    \caption{a) Experimental setup for quantum advantage measurement. It features three parts i) the creation part to transform the polarization entangled state to an ET state ii) the manipulation part consisting of a folded Franson interferometer and iii) the detection part made of filters and detectors. PC : polarization controller. PBS : polarizing beam splitter. BS : beam splitter. WDMs : wavelength-division multiplexers. SNSPD : superconducting nanowire single-photon detectors. TDC : time-to-digital converter. $D_1$ to $D_4$ are the four detectors used for the measurement. The numbers indicated on the WDMs refer to standard telecom wavelength ITU channels. }
    \label{fig: exp}
\end{figure*}
Achieving quantum advantage in photonic interferometry requires maintaining high N-photon interference visibility under realistic imperfections. As an example, defining $\eta$ as the single-photon total transmission and $V$ the interference visibility, sub-SQL measurement must satisfy the inequality $\eta^N V^2 N> 1$~\cite{Slussarenko2017}. For $N=2$, this implies $\eta > 1/\sqrt{2} \approx 70\%$ assuming near-unit visibility, a highly challenging condition to fulfill in realistic scenarios. Moreover, standard implementations such as Franson-type interferometers generate temporally distinguishable components that do not contribute to interference \cite{PhysRevLett.62.2205}. Removing these components typically requires temporal post-selection, which halves the overall detection efficiency, or operating without post-selection, which limits the two-photon fringe visibility to 50\%. As shown analytically in Appendix \ref{Appendix A}, either approach prevents the FI from exceeding the classical bound, making sub-SQL and quantum advantage fundamentally unattainable in such configurations. To overcome this limitation, one method involves the generation of Hong-Ou-Mandel interference at the first beam-splitter (BS) of the interferometer, allowing the photons to coalesce, thus creating a maximally entangled state within the Franson interferometer \cite{10.1063/1.4724105}. However, this perfect coalescence can only take place if the BS is perfectly balanced and the arrival time of the photons at the first BS are equal, leading to a cumbersome experimental apparatus. Another approach, straightforward yet effective, consists of converting a polarization entangled state into an ET state. This method, as well as the experimental apparatus, is depicted in Fig. \ref{fig: exp}. The initial quantum probe of the system is in the state:
\begin{equation}\label{eq: polar entanglement}
    |\psi_{in} \rangle = \frac{|HH\rangle + |VV\rangle}{\sqrt{2}}\, ,
\end{equation}
which describes a maximally entangled two-photon state where $H$ and $V$ denotes the horizontal and vertical polarization, respectively. The photon pairs are generated using a fiber-compatible polarization-entangled source operating at telecom wavelengths, whose design and operating parameters have been carefully optimized to maximize the purity of the generated quantum state \cite{Troisi2026}. We emphasize that the paired photons are wavelength-degenerate around $\lambda = 1560.61 \, \mathrm{nm}$, corresponding to the center of the standard telecom wavelength ITU channel 21. Next, the pairs are sent to an unbalanced interferometer in a Mach-Zehnder configuration, consisting of a polarizing beam splitter (PBS) as the input and a BS at the output. By carefully controlling the polarization of the photons at the input of the PBS by using a polarization controller (PC) to match the state of Eq. \ref{eq: polar entanglement}, the PBS will transcribe the state initially entangled on the polarization basis into a path-entangled state \cite{Kaiser_2013}. Thus, by defining $s$ and $l$ as the bottom and top arms of the interferometer, respectively, the quantum state within the interferometer reads: 
\begin{equation}\label{eq: N00N state theory}
    |\Psi_{in}\rangle = \frac{|20\rangle_{sl} + e^{i\Phi} |02\rangle_{sl}} {\sqrt{2}} \, ,
\end{equation}
where $\Phi$ is the sum of the individual relative phases accumulated by each photon of the entangled state. The latter is materialized in the top arm using a fibered phase shifter. A second PC is placed in the bottom arm to ensure perfect polarization overlap between the two arms. Finally, the pairs are sent in the detection part comprising a series of filters and superconducting nanowire single photon detectors (SNSPD) coupled to a time-to-digital converter (TDC) to register the coincidence events. 

In interferometric measurement, the FI is defined by:
\begin{equation}\label{eq: Fisher info}
    \mathcal{F}_N(\phi)  = \sum_{\substack{i=0 \\ j = N-i}}^{N} \Bigl( \frac{\partial \ln(P_{ij} (\phi))}{\partial \phi}\Bigr)^2 P_{ij} \, ,
\end{equation}
where $P_{ij}$ denotes the different detection probabilities. Thus, its reconstruction requires the knowledge of all the probabilities at the output of the interferometer, \textit{i.e.} when the two photons of the pair exit through the same or two different arms. In the absence of photon-number-resolving detectors, the energy correlations of the paired photons can be exploited to separate them deterministically at the output of the interferometer, as pictured in the detection part of Fig. \ref{fig: exp}. At each output of the interferometer, two cascaded dense wavelength division multiplexers (WDMs) centered at the standard ITU channels 20 and 22 are placed before the detectors. As the paired photons are frequency-degenerate at the center of channel ITU 21 and all the ITU channels have fixed frequency widths (100 GHz), channels 20 and 22 satisfy energy conservation that drives the quantum correlations. This entails that a photon of a pair detected in channel 20 will coincide with its partner in channel 22 and conversely. We label, with respect to the detector's numbers in Fig. \ref{fig: exp}, the output probabilities of the interferometer by $P_{12}$, $P_{34}$, $P_{23}$ and $P_{14}$. Note that the initial polarization entangled pairs are spectrally delocalized in these channels. Consequently, no photons at other frequencies are transmitted through the interferometer, and the experiment's filtering system exclusively guarantees the deterministic separation of photons. \\
As previously stated, accounting for experimental imperfections is mandatory for establishing a rigorous definition of quantum advantage, formally given by:

\begin{equation}\label{eq: QA exp}
\begin{split}
    R & =\frac{1}{2}\frac{Max(\mathcal{F}_2)}{Max(\mathcal{F}_1)}  \\  
    & =2 \frac{V_2^2}{V_1^2}\frac{(\eta_1+\eta_4)(\eta_2+\eta_3)}{\eta_1+\eta_2+\eta_3+\eta_4} > 1 \, ,
\end{split}
\end{equation}
where $\eta_i$ is the global transmission up to the ith detector, including the detector's efficiencies and $V_1$, $V_2$ stand as the single and two-photon fringes visibility, respectively (see Appendix \ref{Appendix B} for calculations). Note that in the case of symmetric global losses $\eta$ and a perfect single and two-photon contrast, the last equation simplifies to:
\begin{equation}
    \eta > \frac{1}{2} \, , 
\end{equation}
highlighting the stringent conditions required to achieve genuine quantum advantage. Consequently, the number of optical components employed in the experiment has been kept to a minimum (without compromising the sensor's potential versatility) and selected with the utmost care to minimize losses. 
\section{Results}

\begin{figure*}[!th]
    \centering
    \includegraphics[width=1\linewidth]{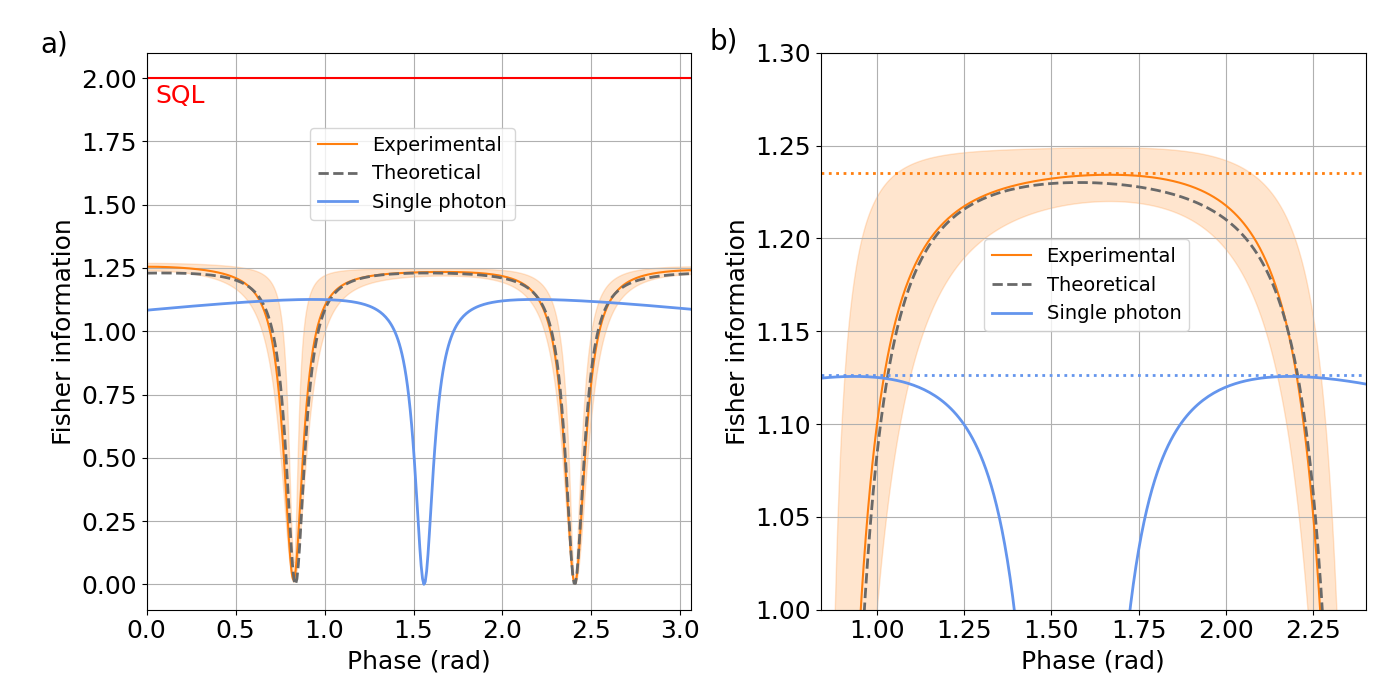}
    \caption{a) FI per photon as a function of the phase extracted from the measured coincidence peaks. The orange curve is the experimental data while the shaded area is the $3\sigma$ (99.7\%) confidence region derived from the fit uncertainty. The blue and grey dashed lines are the theoretical expected $\mathcal{F}_1$ and $\mathcal{F}_2$, respectively. The red line corresponds to the standard quantum limit. b) Zoom in the region of interest. The orange and blue horizontal dotted lines are the maximum of the functions.}
    \label{fig: FI exp}
\end{figure*}

In order to accurately quantify the overall losses of each detection line, the system is tested with a power-calibrated tunable telecom laser whose frequency precisely matches the center of the WDMs employed in the experiment, thereby ensuring a high degree of precision in the losses calibration. 
\begin{figure}[!h]
    \centering
    \includegraphics[width=1\linewidth]{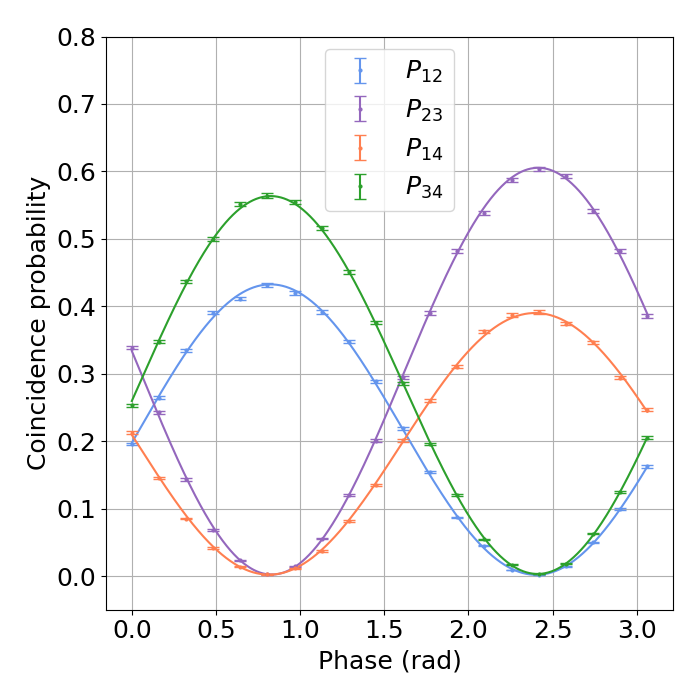}
    \caption{Experimental coincidences measured at the output of the interferometer. The error bars represent the standard Poissonian distribution. The solid lines represent a fit from Eq. \ref{eq: fit vis}.  The difference in amplitude of each of the curves is induced by the asymmetric losses of the system.}
    \label{fig: fringes}
\end{figure}
The same laser is used with a calibrated electronic variable optical attenuator to infer the SNSPD's efficiency. We find for each line the total transmission factors (losses in dB) $\eta_1 = 0.517 \, (2.43), \eta_2 = 0.546 \, (2.63), \eta_3 = 0.649 \, (1.87)$ and $\eta_4 = 0.608 \, (2.16)$. Note that the relative losses between the two arms of the interferometer are $1-\eta_t = 0.12$, where $\eta_t$ is the relative transmission. Assuming that only this parameter affects the interference contrast, the visibilities of single and two-photon interference are estimated to be $V_1 = 99.8 \%$ and $V_2 = 99.2 \%$, respectively (see Appendix \ref{Appendix C} for calculations). From there, we calculate a theoretical quantum advantage $\mathcal{F}_2 / \mathcal{F}_1 \approx 1.09$. Another source of noise that can affect the contrast of the fringes lies in the generation of multiple pairs at the same time. A direct way to verify the absence of the latter is to measure coincidences between two detectors sharing the same ITU filter channels, namely $P_{13}$ and $P_{24}$. Upon acquiring data for several minutes, no coincidence peaks were observed between these detectors, thereby confirming that multi-pair emission is negligible. Finally, the four coincidence peaks, as depicted in Fig. \ref{fig: fringes}, are measured as a function of the voltage applied to the phase shifter. The four normalized experimental curves are then fitted using the standard formula for ET entanglement:
\begin{equation}\label{eq: fit vis}
    P_{ij} = \frac{1}{4}\bigl(1\pm V_2 cos (2\phi)\bigr) \, ,
\end{equation}
from which we find a mean experimental two-photon interference visibility $V_2^{fit} \approx 99.0 \pm 0.2 \, \%$, in good agreement with the expected theoretical value. From the measured global losses and interference fringes, we infer the FI using Eq. \ref{eq: Fisher info}, plotted in Fig. \ref{fig: FI exp} a) together with a zoom over the region where the FI is maximum in Fig. \ref{fig: FI exp} b). The shaded orange area corresponds to the $3\sigma$ (99.7\%) confidence region, calculated from the fit uncertainty. 
The small asymmetry of the FI is induced by the asymmetric losses of the experiment. From the measured global and relative losses and the associated theoretical single-photon visibility $V_1$, we numerically calculate $\mathcal{F}_1$ as a function of the phase, represented by the solid blue line. We then calculate a ratio $R \approx 1.10$ by taking the maximum of the single and two-photon FI (dotted horizontal lines), where $Max(\mathcal{F}_2)\approx1.24$ and $Max(\mathcal{F}_1)\approx1.13$, thereby demonstrating genuine quantum advantage.

These results demonstrate that a genuine quantum advantage can be achieved in a realistic, fully fiber-based interferometer without post-selection, even in the presence of asymmetric losses. The measured 10\% improvement over the classical Fisher-information bound is already sufficient to provide a net metrological gain for a broad class of sensing applications such as temperature, strain, and gyroscopic measurements.
Further reductions of either relative or global losses would naturally extend the applicability of the sensor to higher-loss samples and transducers while preserving the quantum advantage. For example, eliminating relative losses inside the interferometer while maintaining the same overall transmission would increase the ratio $R$ to 1.15. Likewise, replacing the detectors with photon-number-resolving devices of equal efficiency would remove the need for spectral filtering, thereby reducing global losses and potentially increasing this ratio to 1.27. Finally, when placing this work in the broader context of unconditionally beating the shot-noise limit in deployable sensing scenarios, further progress will primarily rely on reducing global losses rather than relative imbalances, whose impact on visibility is comparatively weaker. As a useful benchmark, achieving a total transmission $\eta = 1/\sqrt{2} \approx 70\%$ in a perfectly symmetric configuration (i.e., in the absence of relative losses) would allow the SQL to be reached, corresponding to a Fisher-information ratio of $R=\sqrt{2}$.

\section{Conclusion}
We have demonstrated a genuine quantum advantage in a Franson interferometer by accounting for all system imperfections, in particular overall losses and detector efficiencies. By transcribing a polarization entangled state into an energy-time entangled state, we have been able to circumvent the net losses or the reduction in visibility that would otherwise be induced by the creation of distinguishable states. The energy conservation of the paired photons enabled deterministic filtering, making it possible to detect all the coincidence probabilities at the output of the interferometer, and to infer the experimental Fisher information of the measurement. We believe that the fully fibered nature of the interferometer at telecom wavelengths paves the way for the development of compact and reliable quantum sensors. Beyond this proof-of-concept, the approach is directly compatible with existing telecom infrastructures, opening the way to distributed, network-integrated quantum sensors capable of operating in real-world environments.

\setcounter{section}{0}
\renewcommand{\thesection}{\Alph{section}}

\section{Sub-SQL measurement with distinguishable states}\label{Appendix A}
\begin{figure}[h]
    \centering
    \includegraphics[width=1\linewidth]{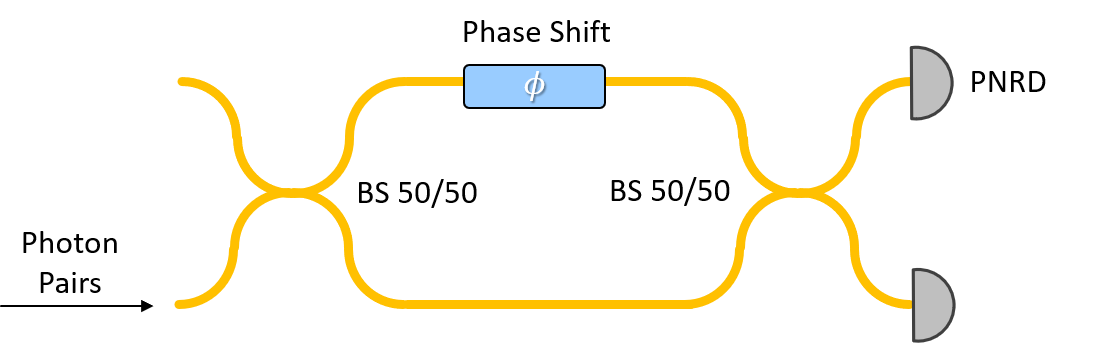}
    \caption{Folded Franson interferometer in a Mach-Zehnder configuration. BS: Beam Splitter; PNRD: Photon Number Resolving Detector.}
    \label{fig: annex a franson}
\end{figure}

We consider a folded Franson interferometer in a Mach-Zehnder configuration described by picture \ref{fig: annex a franson}. The interferometer is arbitrarily unbalanced and the detection system is made of photon-number-resolving detectors with ideal efficiency. Wavelength-degenerated photon pairs enter the system through one arm of the input BS.
Inside the interferometer, the quantum state of the photon pairs reads:

\begin{equation}\label{eq: annex a full noon}
    |\psi\rangle = \frac{1}{2}\left ( |ss\rangle + e^{i2\phi}|ll\rangle + e^{i\phi}(|sl\rangle + |ls\rangle)\right ) \, ,
\end{equation}
where $s$ and $l$ denotes the bottom (short) and top (long) arm of the interferometer, respectively. Provided perfect indistinguishability, the first two superposed states interfere, unlike the latter two. If we temporally post-select the interference term, the global efficiency of the experiment becomes $\eta = 0.5$. If not, the visibility of the interference pattern described by Eq. \ref{eq: fit vis} becomes $V=0.5$. We now calculate the FI of the system by substituting Eq. \ref{eq: fit vis} into Eq. \ref{eq: Fisher info} :

\begin{equation}
\begin{split}
    \mathcal{F}_2(\phi)  & = \sum_{\substack{i=0 \\ j = 2-i}}^{2} \frac{P'^2_{ij}}{P_{ij}} \\
    & =  \left ( \frac{2\eta^2V^2sin^2(2\phi)}{1+Vcos(2\phi)} + \frac{2\eta^2V^2sin^2(2\phi)}{1-Vcos(2\phi)}\right ) \, ,
\end{split}    
\end{equation}
where $P'_{ij}$ denotes the derivative of $P_{ij}$ with respect to $\phi$. The last equation takes its maximum value when $\phi = \frac{\pi}{2}(2k+1)$ where $k\in \mathbf{N}$ and gives:

\begin{equation}
    Max(\mathcal{F}_2) = 4\eta^2V^2 \, ,
\end{equation}
and sub-SQL measurement is defined by :
\begin{equation}\label{eq: annex A superiority}
    4\eta^2V^2 > 2 \, .
\end{equation}
This inequality becomes $\eta > \sqrt{2}$ if the interference term is post-selected or $V > \sqrt{2}$ if it is not. As both inequalities are physically forbidden, sub-SQL measurement cannot be achieved if the quantum state within the Franson interferometer possesses distinguishable components. Moreover, as we have described a perfect system, quantum advantage cannot be achieved as the FI for single photon is maximum $Max(\mathcal{F}_1) =1$ which implies that Eq. \ref{eq: annex A superiority} also describe quantum advantage.

\section{Quantum advantage with asymmetric global losses}\label{Appendix B}
We first calculate the FI for $N=2$ entangled photons, using the coincidence probability given by Eq. \ref{eq: fit vis} and the associated transmission factors:

\begin{equation}
\begin{split}
    \mathcal{F}_2(\phi) & = \sum_{\substack{i=0 \\ j = 2-i}}^{2} \frac{P'^2_{ij}}{P_{ij}} \eta_i\eta_j \\
    & =  2V_2^2sin^2(2\phi) \\
    & \times\left ( \frac{\eta_1\eta_2 + \eta_3 \eta_4}{1+Vcos(2\phi)} + \frac{\eta_2\eta_3 + \eta_1 \eta_4}{1-Vcos(2\phi)}\right ) \, .
\end{split}
\end{equation}
If the asymmetry between the transmission factor of each line is high, a numerical study of the maximum of the FI is needed. If not, the maximum is obtained when $\phi \approx \frac{\pi}{4}(2k+1)$ and gives : 

\begin{equation}
    \begin{split}
        Max(\mathcal{F}_2) & \approx V_2^2(\eta_1\eta_2 + \eta_3 \eta_4 + \eta_2\eta_3 + \eta_1 \eta_4) \\
        & = V_2^2(\eta_1+\eta_3)(\eta_2+\eta_4) \,.
    \end{split}
\end{equation}
For single photon interference, we consider the transmission factors to be $(\eta_1+\eta_2)/2$ and $(\eta_3+\eta_4)/2$. Thus, the detection probability at the output of the interferometer reads:

\begin{equation}
    P_{10,01} = \frac{\eta_{1,3}+\eta_{2,4}}{4} \left ( 1 \pm V_1cos(\phi) \right ) \, ,
\end{equation}
and the FI gives :

\begin{equation}
\begin{split}
    \mathcal{F}_1(\phi) & = \sum_{\substack{i=0 \\ j = 1-i}}^{1} \frac{P'^2_{ij}}{P_{ij}} \frac{(\eta_i+\eta_j)}{2} \\
    & =  \frac{V_1^2}{4}sin^2(\phi) \\
    & \times\left ( \frac{\eta_1+\eta_2}{1+Vcos(\phi)} + \frac{\eta_3+\eta_4}{1-Vcos(\phi)}\right ) \, .
\end{split}
\end{equation}
Again, if the asymmetry between the transmissions factors is low, the maximum of the FI is reached when $\phi \approx \frac{\pi}{2}(2k+1)$ and gives:

\begin{equation}
        Max(\mathcal{F}_1) \approx \frac{V_1^2}{4}(\eta_1+\eta_2 + \eta_3 +\eta_4) \, ,
\end{equation}
from which we derive Eq. \ref{eq: QA exp}.

\section{Interference visibility with relative losses}\label{Appendix C}
We consider the setup depicted in Fig. \ref{fig: annex a franson} where the bottom arm has a perfect transmission and the top arm has a transmission $\eta_t$. We emphasize that $\eta_t$ is expressed in terms of intensity. For sake of simplicity, we consider the state within the interferometer to be a N00N state:

\begin{equation}
    |\psi_N\rangle = \frac{1}{\sqrt{2}} \left ( (\hat{s}^{\dagger})^N+(\sqrt{\eta_t})^Ne^{iN\phi}(\hat{l}^{\dagger})^N   \right )|0\rangle^{\otimes N} \, ,
\end{equation}
where $\hat{s}^{\dagger}$ and $\hat{l}^{\dagger}$ denotes the creation operator in the bottom and top arm of the interferometer, respectively. Thus, the detection probability at the output of the interferometer reads:

\begin{equation}
    P_N = \frac{1}{4} \left (  1 + \eta_t^N \pm  2(\sqrt{\eta_t})^Ncos(N\phi)     \right ) \, ,
\end{equation}
and the associated visibility is:

\begin{equation}
    V_N=\frac{2(\sqrt{\eta_t})^N}{1+\eta_t^N} \, .
\end{equation}
From the last equation, we compute $V_1 = 99.8 \%$ and $V_2 = 99.2 \%$ for $\eta_t = 0.88$.

\section*{Acknowledgment}

This work was conducted within the framework of the OPTIMAL project granted by the European Union by means of the Fond Européen de développement régional (FEDER). The authors also acknowledge financial support from the Agence Nationale de la Recherche (ANR) through the EQUINE project (ANR-23-QUAC-0001), and the French government through its Investments for the Future programme under the Université Côte d'Azur UCA-JEDI project (Quantum@UCA) managed by the ANR (ANR-15-IDEX-01).\\

\section*{Competing interests}
The authors declare that they have no conflict of interest.

\section*{Data Availability}
The data are available from the authors upon request.

\end{document}